\documentstyle[12pt,epsf]{article}
\begin{document}
\begin{titlepage}
    \title{Massive Scalar Particles in a
           Modified Schwarzschild Geometry}
\author{S. Capozziello$^{a,b}$\thanks
{E-mail: capozziello,feoli,lambiase,scarpetta@physics.unisa.it},
 A. Feoli$^{b,c}$, G. Lambiase$^{a,b}$, \\ G. Papini$^{d,e}$\thanks{E-mail:papini@uregina.ca},
 G. Scarpetta$^{a,b,f}$ \\
 {\small{\em $^a$  Dipartimento di Scienze Fisiche ``E.R. Caianiello'',}}\\
 {\small{\em Universit\`a di Salerno, 84081 - Baronissi (SA), Italy.}}  \\
 {\small{\em $^b$ Istituto Nazionale di Fisica Nucleare, Sezione di Napoli}} \\
 {\small{\em $^c$ Facolt\`a di Ingegneria, Universit\`a del Sannio, Benevento, Italy.}}\\
 {\small{\em $^d$ Department of Physics, University of Regina, Regina, Sask. S4S 0A2, Canada.}}\\
 {\small{\em $^e$ Canadian Institute for Theoretical Astrophysics,}}\\
 {\small{\em University of Toronto, Toronto, Ont. K7L 3N6, Canada.}}\\
 {\small{\em $^f$ IIASS, Vietri sul Mare (SA), Italy.}} }

\date{\today}

\maketitle
\begin{abstract}

Massive, spinless bosons have vanishing probability of reaching the sphere $r=2M$ from
the region $r>2M$ when the original Schwarzschild metric is modified by maximal
acceleration corrections.

\end{abstract}

\thispagestyle{empty}  PACS: 04.70.-s, 04.70.Bw

     \vfill
     \end{titlepage}

It is commonly believed that the construction of a geometrical
theory of quantum mechanics would lend perspective to a variety of
problems, from the unification of general relativity and quantum
mechanics to the regularization of field equations. In response to
this need Caianiello and collaborators \cite{qg}, \cite{ma}
developed a model in which quantization is interpreted as
curvature of the eight-dimensional space-time tangent bundle TM.
The model incorporates the Born reciprocity principle and the
notion that the proper acceleration of massive particles has an
upper limit ${\cal A}_m$.

Classical and quantum arguments supporting the existence of a
maximal acceleration (MA) have long been adduced \cite{prove}. MA
also appears in the context of Weyl space \cite{pap} and of a
geometrical analogue of Vigier's stochastic theory \cite{jv}.

Some authors regard ${\cal A}_m$ as a universal constant fixed by
Planck's mass \cite{b},\cite{infl}, but a direct application of
Heisenberg's uncertainty relations \cite{ca},\cite{pw} as well as
the geometrical interpretation of the quantum commutation
relations given by Caianiello, suggest that ${\cal A}_m$ be fixed
by the rest mass of the particle itself according to ${\cal
A}_m=2mc^3/\hbar$.

MA touches upon a number of issues. The existence of a MA would
rid black hole entropy of ultraviolet divergencies
\cite{BHEntropy},\cite{McG}, and circumvent inconsistencies
associated with the application of the point-like concept to
relativistic quantum particles \cite{he}.

It is significant that a limit on the acceleration also occurs in
string theory. Here the upper limit manifests itself through
Jeans-like instabilities \cite{gsv} which occur when the
acceleration induced by the background gravitational field is
larger than a critical value $a_c = (m\alpha)^{-1}$for which the
string extremities become causally disconnected \cite{gasp}. $m$
is the string mass and $\alpha$ is the string tension.

Frolov and Sanchez \cite{fs} have then found that a universal
critical acceleration $a_c = (m\alpha)^{-1}$ must be a general
property of strings.

While in all these instances the critical acceleration is the
result of the interplay of the Rindler horizon with the finite
extension of the particle \cite{emb},\cite{sa2}, in the Caianiello
model MA is a basic physical property of all massive particles
which appears automatically in the physical laws. At the same time
the model introduces an invariant interval in TM that leads to a
regularization of the field equations that does not require a
fundamental length as in \cite{qs} and does therefore preserve the
continuum structure of space-time.

Applications of the Caianiello model range from cosmology to the
calculation of corrections to the Lamb shift of hydrogenic atoms.
A sample of pertinent references can be found in \cite{caian}  .
In all these works space-time is endowed with a causal structure
in which the proper accelerations of massive particles are
limited. This is achieved by means of an embedding procedure
pioneered in \cite{8} and discussed at length in
\cite{emb},\cite{SCH} . The procedure stipulates that the line
element experienced by an accelerating particle is represented by
\begin{equation}\label{eq1}
d\tau^2= \left[1+\frac{\ddot x_\mu\ddot x^\mu}{{\cal
A}_m^2}\right] \eta_{\mu\nu} dx^\mu dx^\nu  {,}
\end{equation}
and is therefore observer-dependent as conjectured by Gibbons and
Hawking \cite{Haw}. As a consequence, the effective space-time
geometry experienced by accelerated particles exhibits
mass-dependent corrections, which in general induce curvature, and
give rise to a mass-dependent violation of the equivalence
principle. The classical limit $\left ({\cal A}_m\right)^{-1} =
{\hbar\over 2 m c^3}\rightarrow 0$ returns space-time to its
ordinary geometry.

In the presence of gravity, we replace $\eta_{\mu\nu}$ with the
corresponding metric tensor $g_{\mu\nu}$, a  choice that preserves
the full structure introduced in the case of flat space. We obtain
 \begin{equation}\label{eq2}
 d\tau^2=\left(1+\frac{g_{\mu\nu}\ddot{x}^{\mu}\ddot{x}^{\nu}}{{\cal A}_m^2}
 \right)g_{\alpha\beta}dx^{\alpha}dx^{\beta}\equiv
 \sigma^2(x) g_{\alpha\beta}dx^{\alpha}dx^{\beta}\,{,}
 \end{equation}
where $\ddot{x}^{\mu}=d^2x^{\mu}/ds^2$ is the, in general,
non--covariant acceleration of a particle along its worldline.

We have recently studied the modifications produced by MA in the
motion of a test particle in a Schwarzschild field \cite{SCH}. We
have found that these account for the presence of a spherical
shell, external to the Schwarzschild sphere, that is forbidden to
any classical particle and hampers the formation of a black hole.
Our aim here is to study the behaviour of a quantum, scalar
particle in this modified Schwarzschild geometry. The calculations
involve both classical and quantum behaviours of the particle
together in a single framework. The first one determines, through
the expectation value of the acceleration, the effective
gravitational field which in turn defines the latter by altering
the make-up of the Klein-Gordon equation.

Before embarking on this problem, some cautionary remarks are in
order \cite{SCH}.

The effective theory presented is intrinsically non-covariant.
Non-covariant is the quadri-acceleration that appears in
$\sigma^2(x)$ and non-covariant is $\sigma^2(x)$ itself which is
not, therefore, a true scalar. In addition $\sigma^2(x)$ could be
eliminated from (\ref{eq2}) by means of a coordinate
transformation if one insisted on applying the principles of
general relativity to this effective theory. On the contrary, the
embedding procedure requires that $\sigma^2(x)$ be present in
(\ref{eq2}) and that it be calculated in the same coordinates of
the unperturbed gravitational background. It is therefore
desirable to check the results of a particular calculation in more
than a single coordinate system. Nonetheless the choice of
$\ddot{x}^{\mu}$ is supported by the derivation of ${\cal A}_m$
from quantum mechanics, by special relativity and by the weak
field approximation to general relativity. A fully covariant
presentation of the ideas expounded is still lacking. The model is
not intended, therefore, to supersede general relativity, but
rather to provide a way to calculate the effect of MA on the
quantum particle.

For convenience, the natural units $\hbar = c = G = 1$ are used
below. The conformal factor can be easily calculated as in
\cite{SCH} starting from (2), with $\theta =\pi/2$, and from the
well known expressions for $\ddot{t},\ddot{r}$ and $\ddot{\phi}$
in Schwarzschild coordinates \cite{wh}. One obtains

$$
\sigma^2(r)=1+\frac{1}{{\cal A}_m^2}\left\{
-\frac{1}{1-2M/r}\left(-\frac{3M\tilde{L}^2}{r^4}+\frac{\tilde{L}^2}{r^3}
-\frac{M}{r^2}\right)^2 + \right.
$$
\begin{equation}\label{1.5}
\left. +\left(-\frac{4\tilde{L}^2}{r^4}+\frac{4\tilde{E}^2
M^2}{r^4(1-2M/r)^3}
\right)\left[\tilde{E}^2-\left(1-\frac{2M}{r}\right)\left(1+
\frac{\tilde{L}^2}{r^2}\right)\right]\right\}\,{,}
\end{equation}
where $M$ is the mass of the source, $\tilde{E}$ and $\tilde{L}$
are the total energy and angular momentum per unit of test
particle rest mass $m$.

As discussed in \cite{SCH}, the modifications introduced by Eq.
(\ref{1.5}) include the presence of a spherical shell, external to
the Schwarzschild sphere, that is forbidden to classical
particles. The radius of the shell is $2M<r<(2+\eta)M$, where
$\eta$ is much less than one and increases with the total energy
per unit of test particle mass $\tilde{E}$. The question now
arises whether quantum particles can penetrate the shell. This
problem is tackled in the present work where the massive, quantum
particle satisfies the Klein--Gordon equation.

In the effective curved space--time of metric (\ref{eq2}),  the
wave equation for a scalar particle of rest mass $m$ is
\begin{equation}\label{weq}
(\nabla_{\mu}\nabla^{\mu} + m^2)\psi(x)=0\,{,}
\end{equation}
where $\nabla_{\mu}\nabla^{\mu}=
(1/\sqrt{-\tilde{g}})\partial_{\mu}
(\sqrt{-\tilde{g}}\tilde{g}^{\mu\nu}
\partial_{\nu})$, $\tilde{g}_{\mu\nu}=\sigma^2(r)g_{\mu\nu}$,
and $\nabla_{\mu}$ is the covariant derivative.

Written explicitly, Eq. (\ref{weq}) takes the form
 $$
 \left\{\frac{\partial^2}{\partial
 t^2}-\frac{e^{\lambda}} {\sigma^2r^2}\frac{\partial} {\partial r}\left(\sigma^2 r^2
 e^{\lambda} \frac{\partial}{\partial r}\right) - \right.
 $$
 \begin{equation}\label{weq1}
\left. - \frac{e^{\lambda}}{r^2}\left[\frac{1}{\sin\theta}\frac{\partial}
{\partial\theta}\left(\sin\theta\frac{\partial}{\partial\theta}\right)+
\frac{1}{\sin^2\theta}\frac{\partial^2}{\partial\phi^2}\right] +
m^2\sigma^2e^{\lambda}\right\}\psi(t,r,\theta,\phi)=0\,{.}
 \end{equation}
By separating variables, the wave function can be
written as
\begin{equation}\label{sep}
\psi(t,r,\theta,\phi) = T(t)R(r)\Theta(\theta,\phi)
\end{equation}
and Eq. (\ref{weq1}) can be split into the  following three equations
\begin{equation}\label{eqT}
\frac{\partial^2 T}{\partial t^2} + \omega^2 T = 0\,{,}
\end{equation}
\begin{equation}\label{eqO}
\frac{1}{\Theta}\left[\frac{1}{\sin\theta}\frac{\partial}{\partial\theta}
\left(\sin\theta\frac{\partial}{\partial\theta}\right)+\frac{1}{\sin^2\theta}
\frac{\partial^2}{\partial\phi^2}\right]\Theta = -l(l+1) \,{,}
\end{equation}
\begin{equation}\label{eqR}
e^{2\lambda}R^{''}+\left(2\frac{\sigma^{\prime}}{\sigma}+
\frac{2}{r}+
\lambda^{\prime}\right)e^{2\lambda}R^{'}+\left[\omega^2-e^{\lambda}
\left(\frac{l(l+1)}{r^2}+m^2\sigma^2\right)\right]R=0\,{,}
\end{equation}
where $\omega^2$ is a separation constant corresponding to the frequency of the wave,
$l$ is the orbital angular momentum quantum number of the scalar particle and a prime
indicates differentiation with respect to $r$.

The solution of Eq. (\ref{eqO}) is
\begin{equation}\label{harm}
\Theta_{lp}(\theta,\phi)=Y^p_l(\cos\theta)e^{ip\phi}
\end{equation}
where $Y^p_l(\cos\theta)$ are the usual spherical harmonics,
and $p$, with $\mid p \mid \leq l$, is the magnetic quantum number.
The general solution of eq. (\ref{eqT}) is
\begin{equation}\label{temp}
T(t)=C_1e^{-i\omega t}+C_2e^{i\omega t}\,{,}
\end{equation}
where $C_1$ and $C_2$ are arbitrary constants. It follows, from
Eqs. (\ref{sep}), (\ref{harm}) and (\ref{temp}) that the
eigenfunctions of the scalar wave equation (\ref{weq1}) can be
cast in the form
\begin{equation}\label{eig}
\psi(t,r,\theta,\phi)=NR(r)Y_l^p(\cos\theta)e^{i(p\phi\pm\omega t)}\,{,}
\end{equation}
where $N$ is a normalization constant and $R(r)$ is the solution of the radial wave
equation (\ref{eqR}).

In order to derive from (\ref{eqR}) the effective quantum
potential in which the boson field propagates, one usually
introduces the variable $r^*=r^*(r)$ such that
\begin{equation}\label{new}
e^{2\lambda}\left(\frac{dr^*}{dr}\right)^2=1\,{.}
\end{equation}
Eq. (\ref{new}) implies that $r^*(r)=r+2M\ln(r-2M)$, which is
defined for $r\geq2M$. After substituting $R(r^*)=\alpha
(r^*)\beta (r^*)$ into (\ref{eqR}), one requires that the
coefficient of $d\alpha/dr^*$ vanishes \cite{Kof}, i.e.
\begin{equation}\label{coef}
\frac{d\beta}{dr^*}-G(r)\beta=0\,{,}
\end{equation}
where
\begin{equation}\label{eqG}
G(r)\equiv -\left(\frac{\sigma^{\prime}}{\sigma}+\frac{1}{r}\right)
e^{\lambda}\,{.}
\end{equation}
In the region $r\geq 2M$, the equation linking $r$ to $r^*$ may be
used to integrate Eq. (\ref{coef}). The result is
$\beta (r)=\beta_0 (r\sigma(r))^{-1}$
where $\beta_0$ is an integration constant. $\beta$ vanishes for $r\to 2M$.
The equation for $\alpha (r)$ reduces to the
Schroedinger--like equation
\begin{equation}\label{sch}
-\frac{d^2\alpha}{dr^{*2}}+V_{eff}(r)\alpha=\omega^2\alpha\,{,}
\end{equation}
where the effective potential $V_{eff}(r)$ is given by
\begin{equation}\label{Veff}
V_{eff}(r)=G^2(r)+e^{\lambda}G^{\prime}(r)+
e^{\lambda}\left(\frac{l(l+1)}{r^2}+m^2\sigma^2\right)\,{.}
\end{equation}
As in \cite{SCH}, it is convenient to introduce the adimensional
quantities $\lambda=\tilde{L}/M= l/(mM)$, $\epsilon=(M{\cal
A}_m)^{-1}=(2mM)^{-1}$ and $\rho=r/M$. The behaviour of
$\tilde{V}_{eff}(\rho)=V_{eff}(\rho)/m^2$ is shown in Fig. 1. The
largest contribution comes from the $e^{\lambda}m^2\sigma^2$ term
in (\ref{sch}). For $\rho\sim 2$ one finds
$\tilde{V}_{eff}(\rho)\sim
\frac{\tilde{E}^4\epsilon^2}{(\rho-2)^2}$, which, unlike
$\tilde{V}_{eff}(\rho)$ of Ref. \cite{SCH}, definitely diverges on
the Schwarzschild sphere. This suggests that $|\alpha |^2 \to 0$
as $r\to 2M$.

\begin{figure}[h]
\epsfysize = 7 cm
\begin{center}
\epsfbox{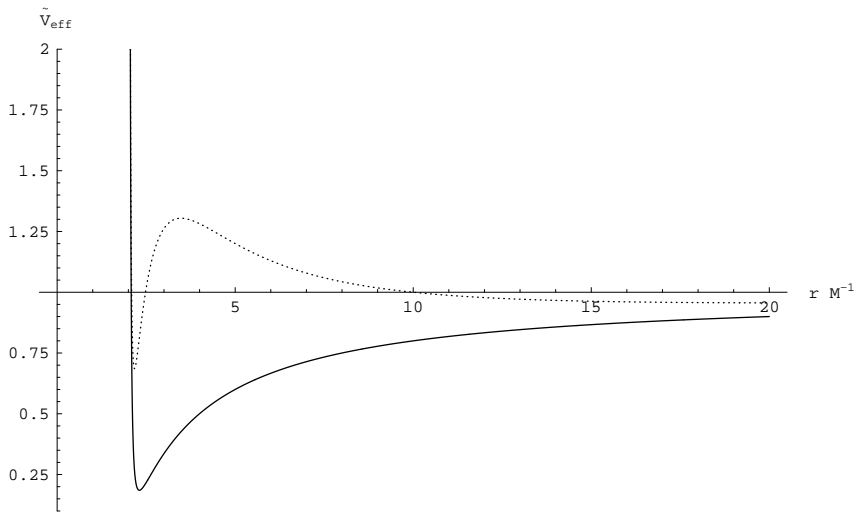}
\caption{\footnotesize {Solid line: effective potential per unit of test particle rest
mass $m$ for $\tilde E = 9$,
 $\epsilon = 0.001 $ and $\lambda = 0$.
Dotted line:  $\tilde E = 9$,
 $\epsilon = 0.001$ and $\lambda = 5$}}
\end{center}
\end{figure}

In order to get a clear indication of
the behaviour of $|R(r)|^2$, we calculate the asymptotic
 solution of the radial wave equation near
the Schwarzschild horizon by writing
\begin{equation}\label{rad}
\rho = 2 + x\,{,}
\end{equation}
where $x<<1$, and by expanding the coefficients of the radial wave equation in a power
series. To leading order, one obtains $\sigma\sim \sqrt{8/\epsilon^2x^3}$ and
$\sigma'/\sigma\sim -3/(2Mx)$. The radial wave equation (\ref{eqR}) then reduces to the
Bessel equation
\begin{equation}\label{near}
x^2\,\frac{d^2R}{dx^2}-2x\,\frac{dR}{dx} - \frac{4}{x^2}\,R=0
\end{equation}
and its solution is
\begin{equation}\label{Bes}
R(x)=x^{3/2}\, e^{\pm i(3/2)\pi}\, Z_{\pm 3/2}\left(\frac{2}{x}\right)\,{,}
\end{equation}
where $Z_{\nu}(z)$ is the Bessel function with half integer index. In the limit $x\to
0$, Eq. (\ref{Bes}) reads
\begin{equation}\label{solu}
R(x)\sim \sqrt{\frac{1}{\pi }} \, x^2\, e^{i(1/x+\gamma)}\,{,}
\end{equation}
where $\gamma$ is a constant phase factor. The radial probability
density $P(x)$ in proximity of the event horizon is then
\begin{equation}\label{pro}
P(x) = \mid R(x) \mid^2 \sim x^4 \,{,}
\end{equation}
which vanishes as $x\to 0$. It must be emphasized that our result
differs from that derived by Kofinti \cite{Kof}. In fact, the
corresponding probability density for propagation of a quantum
particle in an {\it unmodified} Schwarzschild geometry {\it does
not vanish at $\rho=2$} \cite{Kof}.

Let us now ascertain that the results obtained persist in
isotropic coordinates. These are related to $r$ by the non-linear
transformation $r = (1 + a/4u)^2 u$ and yield the metric tensor
\cite{LAN}
\begin{equation}\label{iso1}
  g_{\mu\nu}=\mbox{diag}(e^{\lambda}, -e^{\mu}, -e^{\mu}u^2,
  -e^{\mu}u^2\sin^2\theta),
\end{equation}
where now
 $$
  e^{\lambda}=\frac{(1-a/4u)^2}{(1+a/4u)^2},\quad
  e^{\mu}=\left( 1+\frac{a}{4u}\right)^4,\quad a=2GM/c^2.
  $$
In these coordinates, therefore, the weak field limits of
(\ref{iso1}) and of the Schwarzschild metric coincide. The
isotropic coordinates also leave the element of spatial distance
in conformal form.

We start again from the expressions for the components of the
four-velocity

\begin{eqnarray}
\dot{t}&=&\frac{\tilde{E}(1+a/4u)^2}{(1-a/4u)^2}\,,
\label{eqtis}\\
\dot{u}&=&\frac{1}{(1+a/4u)^2}\left\{\frac{\tilde{E}^2(1+a/4u)^2}{(1-a/4u)^2}-
\frac{\tilde{L}^2}{u^2(1+a/4u)^4}-1\right\}^{1/2}\label{equis}\,,
\\ \dot{\varphi}&=& -\frac{\tilde{L}}{u^2(1+a/4u)^4}
\label{eqphiis}\,.
\end{eqnarray}
and that of the conformal factor
\begin{equation}\label{eq4}
\sigma^2(r)=1+\frac{1}{{\cal A}^2_m}
\left[\frac{(1-a/4u)^2}{(1+a/4u)^2}\ddot{t}^{\,2}-\left(1+\frac{a}{4u}\right)^4
\ddot{u}^2-u^2\left(1+\frac{a}{4u}\right)^4\ddot{\phi}^2\right]\,{.}
\end{equation}
\begin{figure}[h]
\epsfysize = 7 cm
\begin{center}
\epsfbox{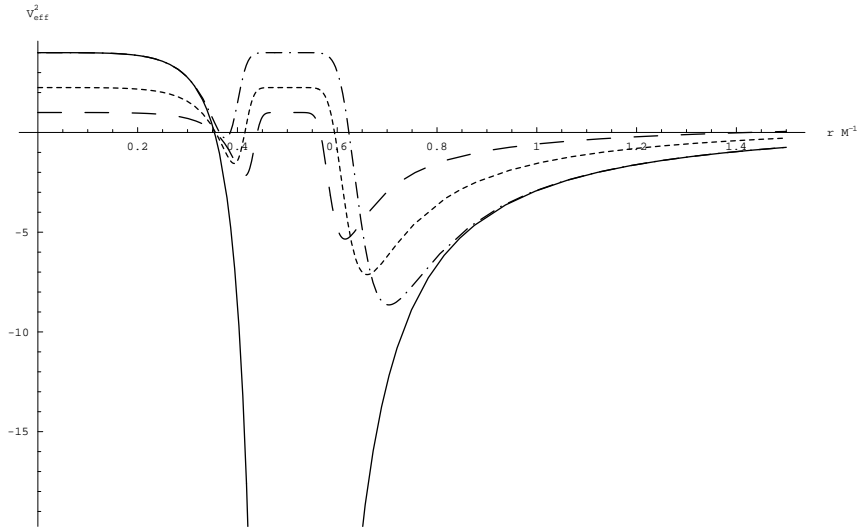} \caption{\footnotesize {Classical potentials in
isotropic coordinates. Solid line: Schwarzschild potential for
 $\epsilon = 0.001 $ and $\lambda = 0$. Dashed line: $\tilde E =
 1$. Dotted line: $\tilde E = 1.5$. Dot-dashed line: $\tilde E =
 2$.}}
\end{center}
\end{figure}
The classical, repulsive shell still exists in proximity of the
horizon $u=a/4$ as indicated by Fig. 2 (compare with Fig. 1 of
Ref. \cite{SCH}). Its existence confirms the result of \cite{SCH}.
The analogous occurrence of a classically impenetrable shell was
also derived by Gasperini \cite{MG} as a consequence of the
breaking of the $SO(3,1)$ symmetry.

 Repeating
the same calculations as in the Schwarzschild case, we find that
the {\it radial} wave function $R(u)$ satisfies the equation
\begin{equation}\label{isoR}
e^{\lambda-\mu}R^{''}+e^{\lambda-\mu}\left(2\frac{\sigma^{\prime}}{\sigma}+
\frac{2}{u}+
\frac{\lambda^{\prime}}{2}+\frac{\mu^{\prime}}{2}\right)R^{'}+
\end{equation}
 $$ +\left[\omega^2-e^{\lambda}
 \left(e^{-\mu}\frac{l(l+1)}{u^2}+m^2\sigma^2\right)\right]R=0\,{.}
 $$
The functions $ \Theta(\theta,\varphi),T(t)$ and the constants are
as defined in (\ref{harm}) and (\ref{eqT}).

In order to derive the effective potential from (\ref{isoR}), we
now introduce the variable $u^*=u^*(u)$ such that
\begin{equation}\label{isonew}
e^{\lambda-\mu}\left(\frac{du^*}{du}\right)^2=1\,{.}
\end{equation}
Eq. (\ref{isonew}) implies that
\begin{equation}\label{isocoord}
 u^*(u)=\frac{a^2}{16}+u-a\ln u+2a\ln(4u-a)\,,
\end{equation}
which is defined for $4u\geq a$. We again substitute
$R(u^*)=\alpha (u^*)\beta (u^*)$ into (\ref{isoR}) and require
that the coefficient of $d\alpha/du^*$ vanishes, i.e.
\begin{equation}\label{isocoef}
\frac{d\beta}{du^*}+G(u)\beta=0\,{,}
\end{equation}
where
\begin{equation}\label{isoeqG}
G(r)\equiv \frac{e^{\lambda/2-\mu/2}}{2}
\left(2\frac{\sigma^{\prime}}{\sigma}+\frac{2}{u}+\mu'\right)
\,{.}
\end{equation}
The integration of (\ref{isocoef}) yields the result
\begin{equation}\label{isobeta}
\beta (u^*)=\frac{\beta_1 e^{-\mu/2}}{\sigma u}\,,
\end{equation}
where $\beta_1$ is an integration constant. The equation for
$\alpha (u^*)$ reduces to the Schroedinger--like equation, where
now the effective potential $V_{eff}(u)$ is given by
\begin{equation}\label{isoVeff}
V_{eff}(u)=\,G^2(u)+e^{\lambda/2-\mu/2}\frac{dG(u)}{du}+
e^{\lambda}\left(e^{-\mu}\,\frac{l(l+1)}{r^2}+m^2\sigma^2\right)\,{.}
\end{equation}
It is a simple task to calculate the behaviour of the potential
(\ref{isoVeff}) in proximity of the singularity point $4u=a$. In
fact, setting $4u=a+x$, with $x\approx 0$, one gets
\begin{equation}\label{isoasymp}
V_{eff}(u)\sim m^2/x^6\to \infty \quad \mbox{as} \quad x\to 0\,,
\end{equation}
where the dominant contribution is represented, once again, by
$e^{\lambda}m^2\sigma^2$. In analogy to the foregoing, we can
therefore conclude that the probability to find the quantum
particle near the horizon vanishes as $u\to a/4$. In fact, in the
limit $x\to 0$, Eq. (\ref{isoR}) reduces to the form ($R\equiv y$)
\begin{equation}\label{isoredR}
  x^2y''-9x y'-\frac{4\tilde{E}^2m^2}{4a^2{\cal A}^2 x^6}y=0\,,
\end{equation}
whose solution is a Bessel function. In the limit $x\to 0$, we get
\begin{equation}\label{isosolR}
  y\sim x^{4}\cos\left(\frac{\tilde{E}m}{2a{\cal A}x^3}\right)\,.
\end{equation}
Then the probability vanishes as $x\to 0$, as expected. The choice
of isotropic coordinates does not alter the fact that massive
scalar particles cannot cross the horizon when propagating in a
space-time modified by MA corrections.

In conclusion, we have determined the behaviour of a spinless
boson in the neighborhood of the Schwarzschild sphere $\rho=2$
when the Schwarzschild metric is modified by maximal acceleration
corrections according to the model of Refs. \cite{qg}, \cite{ma},
\cite{emb}. Though the effective potentials experienced by
classical and quantum particles are different, their effects are
similar. Classical particles can not penetrate the shell of radius
$2<\rho<2+\eta$ where their kinetic energy becomes negative.
Similarly, the probability density to find  massive spinless
bosons in the region $\rho=2+x$ with $x<<1$, vanishes with $x$ at
$\rho=2$ where $\sigma^2(x)$ and the quantum potential diverge. In
both instances, maximal acceleration corrections strongly suppress
the absorption of particles in proximity of the horizon. Quantum
tunneling of scalar particles through the shell is not therefore a
viable process for black hole formation in the model, unless
matter is transmuted first into massless particles, as discussed
in \cite{SCH}. These would then have to be absorbed by the
interior of the star at a rate higher than the corresponding
re--emission rate.

\bigskip
\bigskip

\centerline{\bf Acknowledgments}

\vspace{0.2cm}

Research supported by NATO Collaborative Research Grant No.
970150, by Ministero dell'Universit\`a e della Ricerca Scientifica
of Italy MURST fund 40\% and 60\% art.65 D.P.R. 382/80 and by the
Natural Sciences and Engineering Research Council of Canada. GL
acknowledges the financial support of UE (P.O.M. 1994/1999).

\end{document}